# APPLICATION OF TRANSMISSION BRAGG GRATING FOR VIBRATION MONITORING

Tikhonov E.A., Lyamets A.K.
Institute of Physics National Academy Sci., Kiev, Ukraine, <etikh@iop.kiev.ua>

## 1. INTRODUCTION

In the modern scientific and engineering activity there are fields, which study, analyse, supress or apply various type of vibrations /1/. The broad net of seismographs constantly traces global earthquakes and a tsunami, also seismic signals from nuclear detonations, pressure variations for the ocean bottom called by waves etc. Constantly to be led the international conferences on study of noise and engineering vibrations, solving major problems on safety of functioning of various engineering systems and buildings from bridges and high-rise towers to the large vessels and airplanes. Even the short enumeration of subjects includes following directions of the termed examinations: an aero acoustics, measuring technique of vibrations, a nonlinear acoustics, railway dynamics, automobile and airplane noise and vibrations etc /1-3/.

The diffraction optical structures belong to class the optical elements, carrying out various a tasks of guidance and the analysis of radiation due to the multibeam interference on their own periodic structure. As typical representative of similar elements it is possible to consider the volume periodic structure that has been realized with holographic methods: holographic phase gratings (HPG) /4,5,6/.

Used in given work transmission HPG with the efficient (up to 100%) Bragg diffraction are recorded in real time by argon laser on the self-developing photopolymer PPC-488 that has been developed and used in author's institute /5/. Bragg character of diffraction of a transmitted radiation on such grating is accompanied by 2-beam formation on output - the zero and first order of diffraction, besides diffraction process is efficient in quite narrow gap of incident angles concerning an exact Bragg value: from tens of minutes to several degrees - depending on third dimension of a grating - its thickness. 3D character of such gratings makes its similar to a photon crystal and simultenesly improves its spectral resolving ability /6,8/.

The first study of specific application the transmission HPG for monitoring of small low frequency vibrations was the purpose of the present work. In the given contribution the results confirming the possibility and peculiarities of application of the optical system on the basis HPG and a laser for vibration monitoring are presented.

## 2. QUALITATIVE MODEL OF A GRATING VIBRATION SENSOR

For the quantitative description of light diffraction on HPG the theoretical model of the coupled waves was developed and successfully applied. The quite successful model was based on approximation 2-coupled wave interaction (transmitted and diffracted) in periodic structure with the harmonious space index modulation /7/. The degree adequacy the given model to considered photopolymer gratings is not marginal and needs demonstrations which have been provided in our work /6/. Measuring of a space period of manufactured HPG is not related to the specified theory and grounded on marginal usability of the Bragg condition to monochromic light diffraction on the similar periodic structure.

Cook-Klein's measure of reference of a light diffraction on the considered gratings to multiwave or 2 wave is another important component for the theory /7/. Exact estimates behind this measure are possible at direct method of measurement of an actual thickness T and a spatial period of the grating $\Lambda$ or by analysis on the basis of actual measuring of their angular selectivity formfactor.

The short mentioned presentations and the stored experience of photopolymer HPG manufacture allows us to record the transmission Bragg gratings with an angular selectivity from some degrees to some tens of minutes purposefully. An estimate of spectral resolving ability and application a such HPG for measuring of optical wavelengths in a visible spectral band was presented in our work /8/.

The essence of proposed application of volume HPG for vibration registration is conjugated with presence of the sharp angular dependence diffraction efficiency (DE) around Bragg angle for laser beams, i.e. beams with small divergence. Therefore angular diversions in system HPG - a wave vector of an incident beam lead to an anti-phase amplitude modulation of output radiation powers for zero and first diffraction order beams that can be registered by correspondent electronics tools. Initial estimates of expected sensitivity of a such system to angular detuning can be gained at knowledge of angular selectivity volume HPG and a divergence of the laser beam. If angular selectivity of HPG is over the range (20-60) angular minits and used laser beam has typical divergence $\approx$1mrad, so it is possible to detect out angles of displacement not less than the laser bream divergency.

For comparison let's give the threshold sensitivity of the modern standard seismographs of type APCC (designs 2009 year from institute of Oceanographic technique of the Russian Academy of Sciences) on frequency of vibrations 1Hz makes $(10^{-7}-10^{-8})$ km/s. If the linear biases matching to given sensitivity can call the laser beam displacement on (3-4) minutes the proposed system HPG+laser can register the angular displacement at use of the shoulder exceeding length $(10^{-7}-10^{-8})$ m / $(10^{-4}$ rad$) = (0,1-1)$mm that is simply carry out in construction. Accordingly, the similar system on the basis of diode injection lasers and the volume HPG can be considered as a sensor of vibrations of the arbitrary nature calling angular detuning in dispersion plane of grating.

On the other hand, the study and determination of the requirements setting best sensitivity of the volume HPG to vibrations can be useful at solution an inverse task of minimization of the termed sensitivity at making on the basis of such gratings the optical systems of other appointment: interferometers, polychromators, adjustable attenuators of light, etc.

### 3. ELEMENTARY ANALYSIS OF HPG + LASER VIBRATION SENSOR

The angular dependences of diffraction efficiency for s, p- linear polarized waves for considered HPG gained in /7/ it is possible to present as the following expressions convenient for the analysis form /6/:

$$\eta_s = \sin^2 v_s [(1+(\frac{n}{n_1}\delta \sin(2\theta))^2]^{0,5} /[(1+(\frac{n}{n_1}\delta \sin(2\theta))^2] \quad (1a)$$

$$\eta_p = \sin^2 v_p [(1+(\frac{n}{n_1}\delta \tan(2\theta))^2]^{0,5} /[(1+(\frac{n}{n_1}\delta \tan(2\theta))^2] \quad (1б)$$

$$v_s = \frac{\pi \cdot n_1 \cdot T}{\lambda \cos\theta}, \quad v_p = \frac{\pi \cdot n_1 \cdot T \cdot \cos(2\cdot\theta)}{\lambda \cdot \cos\theta}, \quad (1c)$$

here $\theta$, $\lambda$ - the interior Bragg angle at the zero detuning and diffraction wavelength, $\delta = |\Theta_0 - \Theta|$ -angular detuning concerning an exact Bragg angle, T - the thickness and $\Lambda$ - the space grating period, n and $n_1$ - refractive index of photopolymer and amplitude of its space modulation resulting after an holographic grating record. The given dependences specify that DE=max is determined only by grating "force" $\nu$, and the formfactor of angular dependence is determined by sink-function in square brackets. It allows to control in necessary limits the DE formfactor in a recorded grating and its dependence on a transmitted light wavelength.

In fig.1. typical dependence of DE on a detuning concerning an exact Bragg angle $\theta=41,25^0$ for a s-polarisation wave 632,8nm He-Ne - laser is presented. It is remarkably featured by first of the above-stated design formulas (1a). This HPG has been used at experiments with an angular diversion of an incident beam under artificial vibrations. By the way of illustration rated dependence of DE of a HPG and its derivative on a detuning angle $\delta= \Theta_0 - \Theta$. is presented also on fig.2a,b.

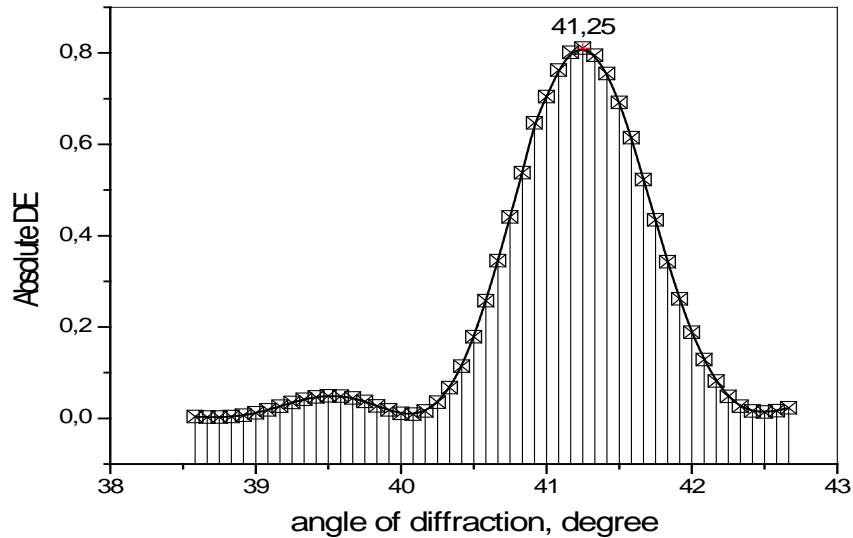

**Fig. 1**. Measured DE of transmission HPG on a wavelength 632,8nm, s-orientation of linear polarized light

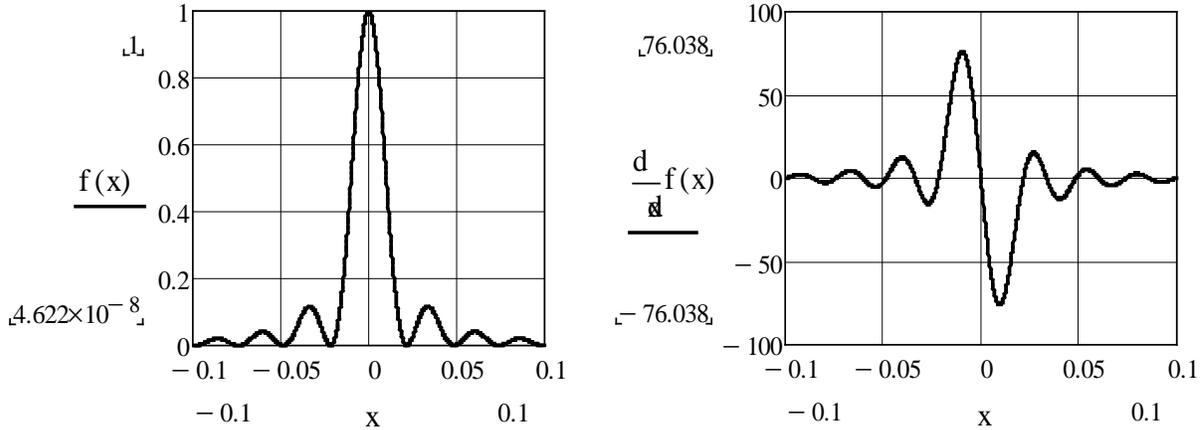

**Fig. 2a.b.** Design values of DE and derivative of DE on an angular detuning (x - in radians) at phase shift («force of grating»), equal π/2 .

It is possible to see that at an angular diversion in a plane of diffraction HPG of a laser beam with divergence a lot of smaller of grating angular selectivity will occur an antiphase amplitude modulation of power of transiting and diffracted beams.

For the subsequent analysis of modulation of diffracting radiation on a vibrating grating we will spot a standing of the characteristic operating point on curve DE and its derivative on fig. 2 and 3. These are fields of maximum DE=max and a zero value of a derivative of DE on a detuning: $\partial ДЭ / \partial \delta$ =0 on fig. 2a,b accordingly. Operating areas on falling left and right parts of dependence of DE coincide with a standing of main maximums of a derivative of DE on a detuning also are visible on specified fig. 2a,b.

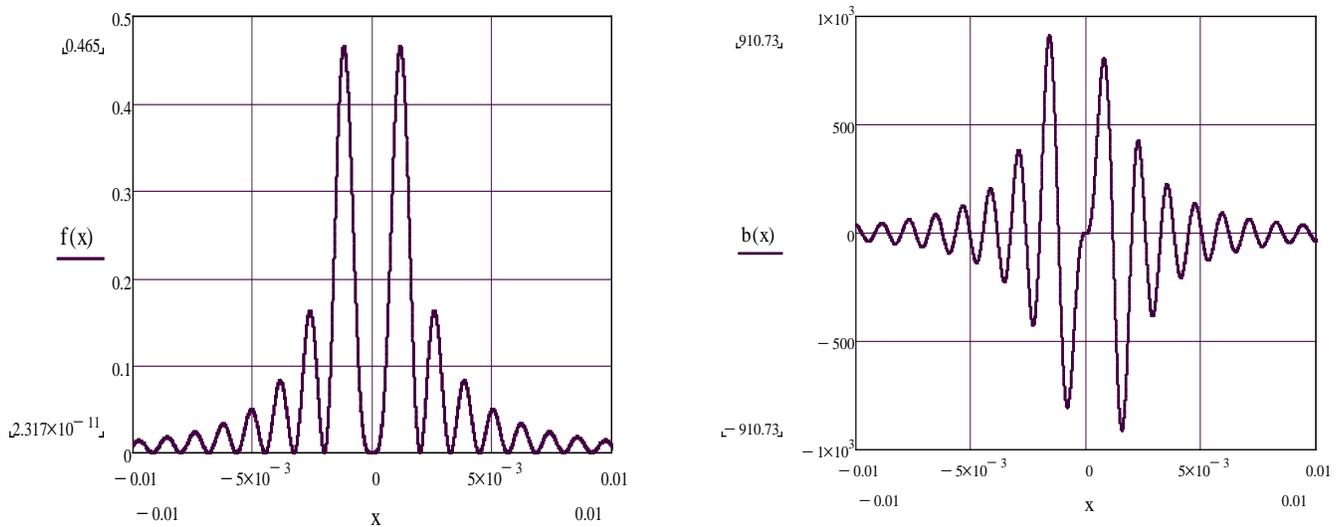

Fig. 3a,b. Angular dependence of DE and its derivative on an angular detuning at phase shift equal π (x - in radians).

In case of phase progression equal π, rated dependence of DE for a similar grating is presented on fig. 3a. In this case DE has a zero value on exact frequency of the Bragg and a peak value in symmetrically arranged 2 maximums.

Let's guess that the angular detuning between HPG and beam (with divergence small in comparison with angular selectivity of the grating defined by a standing of the first nulls on dependence DE=f ($\delta$)) occurs. Let variations of a detuning concerning an operating point $\theta'$ submit to the harmonious approximation $\delta(t) = \delta_0 \sin(\Omega t)$. In the determined conditions it is possible to present dependence of diffracted power by a Taylor series with values of derivatives on a diffraction angle $\theta$ in an operating point $\theta'$:

$$P_{out}(t) = P_0 \eta(\theta' \pm \delta_0 \sin(\Omega t)) \simeq P_0 [\eta(\theta') \pm \delta_0 \sin(\Omega t)(\frac{d\eta(\theta)}{d\theta})_{\theta'} \pm \frac{1}{2!}(\delta_0 \sin(\Omega t))^2 (\frac{d^2\eta(\theta)}{d\theta^2})_{\theta'} \pm ...] \quad (2)$$

Depending on an operating point select on an angle $\theta'$, DE($\theta$) and its derivatives will have significant or zero values, and it will effect on a relation between frequencies of vibrations $\Omega$ on HFG input and on an exit after diffraction. For example, if at $\theta=\theta_0$, the detuning is absnent, DE=max, the first derivative of DE on an angle is equal to zero, but second one accepts a peak value, providing a modification of going out diffracted power on double frequency 2 $\Omega$. In case of an operating point select in a position matching to 2 symmetrical maximums of a derivative (fig. 2б), modulation P (t) happens on frequency of vibrations forcing it. These and other possibilities will be observationally illustrated.

Other possibilities originate at similar angular scanning of a grating (or input beam) in case of HPG, phase shift for which is equal $\pi$ (fig.3a). At a select of the operating point, satisfying to the strict Bragg requirement, the direct emission component in a going out diffracted power misses at all as DE($\theta_0$)=0, while the alternative component shows emission on the doubled scan frequency for the same parents, as in the first featured case. The specified angle dependence of DE can be used in systems of beam stabilization in space.

## 4. EXPERIMANTAL RESULTS

The approbation of a sensor of vibrations was carried out in the optical scheme constituted from HPG with different magnitudes of angular selectivity, He-Ne laser on 632,8nm with beam divergence 1-2mrad and s-orientations of its polarization plane. The vibrating mirror was interfaced to the electromechanical actuator operating at a frequency $\approx$160 Hz and giving rise to angular scan in a grating dispersion plane with amplitude to 30 min. The grating were set on a goniometric table for their smooth angular turn and a select of an initial angle of incidence - an initial operating point concerning which angular scanning by vibrations was carried out. The diffracted beam and a beam of zero diffraction order directed on semiconductor detectors PhD-24 working in the photodiode regime which were switched on inputs of the 2-beam oscilloscope.

At an operating point sampling on one of 2 falling branches concerning a maximum DE scanning of an input beam on frequency $\Omega$ was accompanied by expected antiphase modulation of the diffracted and feed-through emissions on the same frequency (fig. 4).

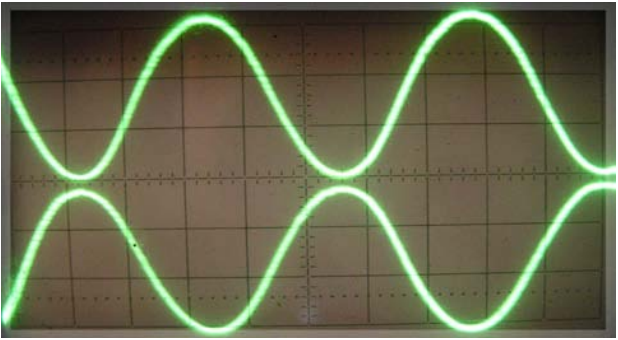

Fig. 4. Antiphase modulation of beam power on a grating exit at angular scanning of an input beam with amplitude some angular minutes. The operating point is set in a maximum of a DE derivative on a diffraction angle.

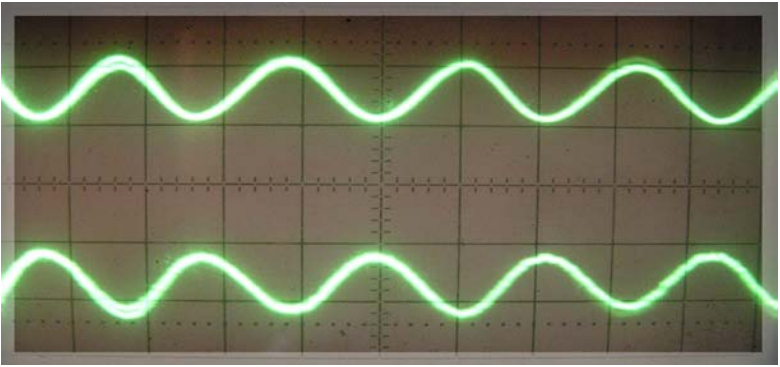

Fig. 5. The operating point coincides with a DE maximum and a maximum of a second DE derivative. The scanning frequency is the former as on fig. 4., but observed modulation frequency of output radiation is doubling.

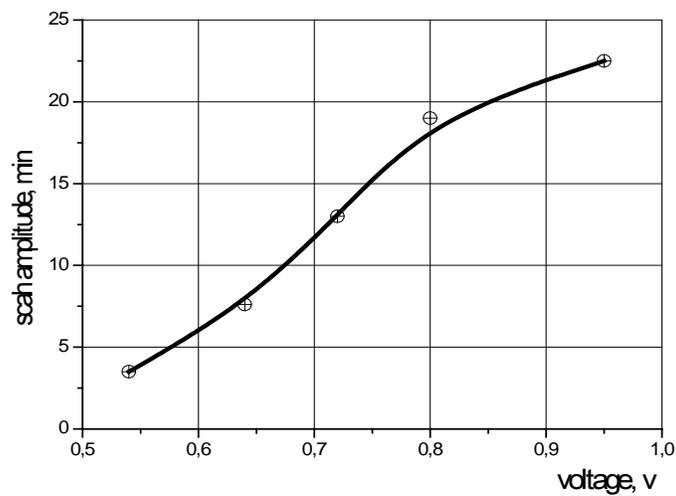

Fig. 6. Dependence of amplitude of the angular scanning of a beam from an applied voltage on the used actuator

At replacement of the used grating with angular selectivity $2,4^0$ with one at selectivity $5^0$ (at other invariable parameters) the modulation depth decreased multiply. The peak amplitude of scanning in the given instances did not exceed one degree and matched to angular selectivity of the chosen grating ( fig. 1.). It was seen that the amplitude of the modulated signal multiply exceeded a noise level of registering system at radiated power of the laser nearby 10мВт.

The minimum scanning angle registered by the tested sensor by supply voltage decrease to the vibration actuator has been measured. Results of measuring are presented on fig. 6. Showed that at scanning amplitude close to angular beam divergence of the used laser the pickup signal in the form of the modulated diffracted powerl dicreased to a zero value.

## 5. CONCLUSIONS

- ✓ It is shown that proposed optical-electronic system in composition of the volume transmission grating, a laser and intermediate sensitive to vibrations mirror detects vibrations when the angular scanning amplitude of an incident beam exceeds angular divergence of a laser beam provided that initial diffraction angle coinciding with a maximum of the first DE angular derivative of the used grating.
- ✓ Sensitivity of sensory system to vibrations in angular minutes depends on an operating point sampling on angular DE dependence and takes max value for grating with angular selectivity comparable with angular divergence of the max laser beam quality.
- ✓ Mathematical model of functioning of a sensor in the form of a Taylor series well featuring a sensitivity it and effect of a response frequency doubling to relative frequency of vibration is presented.

## 6. THE QUOTED LITERATURE